\documentstyle[aps,prl,floats]{revtex}

\begin{document}


\draft
\twocolumn[\hsize\textwidth\columnwidth\hsize\csname @twocolumnfalse\endcsname
%
\title{Stable Spin Precession at one Half of Equilibrium Magnetization in
Superfluid $^3$He-B}
\author{V.V. Dmitriev$^{1,2}$, I.V. Kosarev$^1$,  M. Krusius$^2$, D.V.
Ponarin$^1$,  V.M.H. Ruutu$^2$, and G.E. Volovik$^{2,3}$}

\address{$^1$Kapitza Institute for Physical Problems, 117334
Moscow, Russia, \\$^2$Low Temperature Laboratory, Helsinki University of
Technology, 02150 Espoo, Finland,\\
$^3$Landau Institute for Theoretical Physics, 117334
Moscow, Russia.}

\date{\today} \maketitle

\begin{abstract}
New stable modes of spin precession have been observed in superfluid $^3$He-B.
These dynamical order parameter states include  precession with a magnetization
$S=pS_{eq}$ which is different from the  equilibrium value $S_{eq}$. We have
identified modes with $p=1$, 1/2 and $\approx 0$. The $p= 1/2$ mode is the
second member of phase correlated states of a spin superfluid. The new
states can
be excited in the temperature range $1-T/T_c \lesssim 0.02$ where the energy
barriers between the different local minima of the spin-orbit energy are small.
They are stable in CW NMR due to low dissipation close to $T_c$.

\end{abstract}

\bigskip
] 

A number of stable states of spin precession have been observed and
identified in
NMR experiments on the normal and B phases of liquid $^3$He and on
$^3$He-$^4$He solutions \cite{Borovik-Fomin,Bunkov-Fisher2,Nunes,Dmitriev}.
Some of these states display the same features of spontaneously broken
symmetry, which characterize quantum coherence in superfluids and
superconductors in general. Here the role of the supercurrent is played by spin
currents \cite{HPD-review}. The first representative of such
states -- the Homogeneously Precessing Domain (HPD) in $^3$He-B -- is
clearly phase coherent since its spin-current displays the Josephson effect,
it supports topological spin-current vortices and collective modes, which result
from the spin rigidity of this dynamical state. The main difference from the
more generic  superfluid and superconducting systems is the following: The mass
current conserves the particle number while in the coherently precessing
states the
spin is not a conserved quantity due to the spin-orbit (dipole)
interaction. To stabilize a steady state of precession external
radio-frequency (RF) pumping has to be applied to compensate for spin and
energy relaxation. The smaller the dissipation, the smaller can be the
applied RF
field, and the closer the system will be to the limit of an ideal coherent
dynamical state of a spin superfluid.

We report the observation of new precessing states in $^3$He-B with nonlinear
dynamics of the order parameter. These states were stabilized with
continuous wave (CW) NMR in the vicinity of the superfluid transition
temperature $T_c$.  As distinct from the HPD mode, where the
precessing magnetization has the thermal equilibrium value
$S_{\rm eq}=(\chi/\gamma) H$ ($\chi$ is the susceptibility,
$\gamma$ the gyromagnetic ratio, and $H$ the static external magnetic field),
some of the new states have a fractional value $S=pS_{\rm eq}$,
with $p\approx 0.5$, in particular. The new modes have been identified by
direct
measurement of the magnetization using a combination of  CW and pulsed NMR
techniques. At fixed $H$ in the desired precessing state we switched off the
CW RF pumping. Immediately after that a $90^\circ$ tipping pulse was
applied and the free induction decay (FID) signal was recorded (inset in
Fig.~1). On comparing its amplitude with that of the FID obtained after a
$90^\circ$ tipping pulse in normal $^3$He just
above $T_c$ we find that for all  {\it half-magnetization} (HM) states the
amplitude ratio is $0.50\pm 0.03$ and not dependent on the initial phase of
the tipping pulse. The latter means that the magnetization is oriented
mostly along ${\bf H}$. The amplitude of the  FID signal then indicates that
$S={1\over 2}S_{eq}$. The duration of the FID in the HM
states is closely the same as in normal $^3$He. This implies that the  HM
states preserve their configuration even in the absence of RF pumping.

{\it  Experiments}.--Measurements were carried out with nuclear demagnetization
cryostats both in Helsinki and in Moscow. In Helsinki a cylindrical quartz NMR
cell with a diameter $D=5$ mm and length {\it L}=7 mm was connected to the
rest of
the $^3$He chamber via a small orifice in the bottom of the cell. In Moscow a
similar set-up was used with $D=3.7$ mm and {\it L}=4 mm. The
inside surface of this second container was covered with lavsan (optically
smooth  polyethylen foil similar to mylar). The measurements were
performed in   fields $ H = 73$ --- 284 Oe (corresponding to the
Larmor  frequencies $\omega_L/2\pi =\gamma H/2\pi $ of 235 to 920
kHz), oriented parallel to  the container axis at pressures
$P=0$ --- 12 bar. In both cryostats we used standard CW NMR
spectrometers to monitor the in-phase dispersion ($\propto
S_x$) and out-of-phase absorption ($\propto S_y$) signals. In Moscow pulsed
NMR measurements were also performed and, to check the influence of spatial
inhomogeneity on the spin dynamics,  the  homogeneity of the
$H$ field was improved from $\delta H/H \sim 7\cdot10^{-4}$ in Helsinki to
$5\cdot10^{-5}$. However, it was found that the application of a
small gradient in the axial orientation did not change the
results qualitatively.

In both cryostats we found that in large transverse  RF field
$H_{\rm RF}\sim 0.02-0.06$ Oe  the CW NMR line shape  drastically changes at
temperatures  $T\gtrsim 0.98\,T_c$. In  Fig.~1 the absorption and
dispersion signals are shown, measured at fixed excitation frequency
$\omega_{RF}$  during a slow linear sweep of the static field $H$. The
line-shape
displays abrupt discontinuities, which are reproducible from one
measurement to the next and indicate transitions between different
states of spin precession. The transitions appear at negative frequency shift
$\omega_{\rm RF} - \gamma H$ during a sweep in the direction
of increasing field (Fig.~1) or at positive frequency shift during a sweep down
(Fig.~2). The transitions are mostly of first order since they are  hysteretic
with respect to the direction of the field sweep (dashed line in Fig.~2). The
line shapes in Fig.~2 have been recorded during a slow warm-up. The lower most
spectrum is outside the temperature region of the new states and represents
the usual  NMR line with the excitation of spin waves in the static ${\bf {\hat
n}}$ texture at low RF level \cite{VollhardtWolfle}. We call it here the {\it
spin wave mode} (SW). At higher temperatures transitions to new states appear,
such that a sequence of different states is traversed during the field sweep as
the temperature increases towards $T_c$.

{
\renewcommand{\baselinestretch}{1.0}
\small
\begin{figure}[!!!t]
  \begin{center}
    \leavevmode
\caption{
{\it Inset}: Free induction decay signals after a $90^\circ$ tipping
pulse from the HM4 mode (thick lines)  and from normal $^3$He (thin lines). The
ratio of the amplitudes is $0.52$ initially. The signals are recordered with a
digital oscilloscope. {\it Main frame:} Measured absorption ($S_y$) and
dispersion
($S_x$)  during a sweep of increasing field. (a) Lavsan cell, $P=0$, $T\sim
0.997~T_c$,
$H_{\rm RF}\sim 0.02$ Oe, $\omega_{\rm RF}/2\pi = 334$ kHz.
(b) Quartz cell, $P=5$ bar, $T\sim 0.99~T_c$,
$H_{\rm RF}=0.038$ Oe; $\omega_{\rm RF}/2\pi = 380$ kHz. The FID in the
inset was measured at the value of $H$ marked by the open arrow.}
  \label{f.1}
 \end{center}
\end{figure}
}
{\it Precessing States.}--To identify the different states in Figs.~1
and 2 we determine with the help of the NMR
spectrum the dependence of the total magnetization $S$  and its transverse
component $S_\perp= S \sin\beta = \sqrt{S_x^2+S_y^2}$ on $T$, $H_{\rm RF}$
and $\omega_{\rm RF}-\omega_L$, and compare with analytical or
numerical results. A number of stable solutions have been proposed
to Leggett's equations of $^3$He-B spin dynamics
\cite{Smith-Brinkman,Sonin,Kharadze-Vachnadze,Volovik}. The relevant cases
have been summarized in Fig.~3 in up-dated form. In our numerical
simulation of the  CW NMR measurement we used a spatially homogeneous $H$, with
otherwise realistic values of the experimental parameters, the dipole
interaction
and Leggett-Takagi relaxation in the bulk liquid
\cite{VollhardtWolfle}.

{
\renewcommand{\baselinestretch}{1.0}
\small
\begin{figure}[!!!t]
  \begin{center}
    \leavevmode
\caption{
Dispersion ($S_x$) signal
during downward field sweep and slow warm-up:  Lavsan cell, $P=0$,
$H_{\rm RF}\sim 0.02$ Oe; $\omega_{\rm RF}/2\pi = 461$ kHz. The dashed line
corresponds to an upward field sweep starting from the SW mode. The modes marked
as A appear in the region close to the Larmor field where the remanent
inhomogeneity complicates identification.  }
  \label{f.2}
 \end{center}
\end{figure}
}
In Fig.~3 the different states are classified in terms of the magnitude and
direction of the spin ${\bf S}$ in the precessing frame, and the orientation of
the order parameter, which in $^3$He-B can be represented by the direction
of the
density of the orbital momentum of Cooper pairs ${\bf L}$ in the laboratory
frame.
For the isotropic $J=0$ state of $^3$He-B the magnitude of the orbital momentum
$L=S$. The existence of many different stable states is due to
the particular form of the spin-orbit interaction $F_D$. If the Zeeman energy
is much larger than $F_D$, then $F_D$ is averaged over the
fast Larmor precession and its average  $F_D(s_z,l_z)$ depends on the
orientations of ${\bf S}$ and ${\bf L}$. Here $s_z=S_z/S$,
$l_z=L_z/L$, and ${\bf \hat z}$ is the direction of ${\bf H}$. $F_D$ has a
local minimum in the so-called resonance case, when the magnitude of
$S\approx S_{\rm eq}$ \cite{Fomin1978}. Two more resonance cases with local
minima
of $F_D$  have been predicted to exist \cite{Kharadze-Vachnadze},
namely with  $S\approx pS_{\rm eq}$, where $p=1/2$ or 2. The stable precessing
states are concentrated near these main attractors in ${\bf S}-{\bf L}$ space.

{
\renewcommand{\baselinestretch}{1.0}
\small
\begin{figure}[!!!t]
  \begin{center}
    \leavevmode
\caption{Diagram of precessing states in the plane $S_z/S_{\rm eq}$ --
$S_\perp/S_{\rm eq} $, where ${\bf S}_{\rm eq}=(\chi/\gamma){\bf H}$. SW
($\Box$) denotes  conventional NMR where spin waves are excited in the
stationary order parameter texture. The BS mode exists on the semicircle
$S=S_{\rm
eq}$ (thick black line), with ${\bf L} \parallel \bf H$. The phase-coherent
HPD  state  is the dashed part of the BS
semicircle where $\beta> 104^\circ$. The BS mode with $\beta<104^\circ$ is
stabilized for the first time in the present CW NMR experiments. The HM states
(dashed semicircle) have ${1 \over 2} S_{\rm eq}$ while nearly
zero magnetization is observed in the ZM states. The observed ZM
($\triangle$) and
HM ($\bigcirc$) states are close to the line of aligned spins (AS) with
${\bf S} \parallel {\bf H} \perp {\bf L}$.}
  \label{f.3}
 \end{center}
\end{figure}
}
The resonance phenomenon can be pictured as double resonance. In the
frame rotating about the direction  of  ${\bf H}$ with the
frequency $\omega_{\rm RF} $, the RF field $H_{\rm RF}$ is constant
and the total uncompensated field is:
$$\tilde{\bf H}= {\bf {\hat z}} (\omega_L-\omega_{\rm RF})/ \gamma   + {\bf
{\hat
x}}  H_{\rm RF}~~.\eqno(1)$$
The doubly precessing states are excited by the dipole torque, which
generates the resonance harmonic $\tilde\omega_L = \gamma {\tilde H}$,  if the
following condition is satisfied by the magnitude of the magnetization:
$$S=(\chi/\gamma^2) (~p~ \omega_{\rm RF}+ \gamma\tilde H)~~, \eqno(2)$$
where $p=0,1/2,1$, or 2. In the experiment magnetic relaxation and spatial
inhomogeneity are important,  and Eq.~(2) becomes approximate, but the nature of
the states remains unchanged.

{\it Brinkman-Smith Mode}.--Conventional linear CW  NMR response in the SW mode
corresponds to small oscillations of ${\bf S}$ about  ${\bf S}_{\rm eq}$
(Fig.~3).  In this  state the orienting effect of the field ${\bf H}$ on
${\bf L}$
is weak and spatially the
${\bf L}$-vector  forms a broad texture due to the boundary conditions
\cite{VollhardtWolfle}. When the RF amplitude is increased, the SW mode
becomes unstable  and generally the HPD mode takes over. Here spin currents
maintain phase coherence of the precession of ${\bf S}$ at an angle
$\beta \gtrsim \theta_L$,  where $\theta_L \approx 104^\circ$ is the
Leggett angle
\cite{VollhardtWolfle}. The precursor of the HPD, the {\it Brinkman-Smith} mode
(BS) \cite{Smith-Brinkman} was actually the first of the extraodinary precessing
states which was  observed in early pulsed NMR experiments \cite{Corruccini}.
We find now that the BS mode with $\beta<\theta_L$
can also be stabilized in CW NMR in our quartz cell (Fig.~1). In this case
${\bf S}$  is oriented along $\tilde{\bf H}$ in the rotating frame
\cite{Volovik},
$${\bf S}_{\rm BS}\approx S_{eq} \tilde{\bf H}/ \tilde H
~~,~~\sin\beta=  H_{\rm RF}/ \tilde H ~.
\eqno(3)$$
The BS mode is identified by its transverse spin $S_\perp = S \sin \beta$, which
was found to be proportional to the RF level and inversely
proportional to the frequency shift, with no  dependence on $T$, in agreement
with Eq.~3.

In the BS state the dipole energy orients ${\bf L}$ along ${\bf H}$  (Fig.~3).
This leads to textural
transitions, which explain the discontinuities in Fig.~1 on moving from the SW
into the BS state and back. Such a textural transition was observed
in pulsed NMR \cite{BorovikBunkovDmitriev}, as suggested in
Ref.~\cite{GoloLemanFomin}.

{\it Half-Magnetization Modes}.--We observe four different HM states. HM1 -- HM3
exist at $\omega_{\rm RF} > \omega_L$  and differ from each other by the
sign and
magnitude of the dispersion (Fig.~2). HM4 exists only at $\omega_{\rm
RF} <\omega_L$ (Fig.~1).  All these states lie in Fig.~3 in the vicinity of the
positive intersection of the $p=1/2$ circle with the vertical line, the
locus of the aligned spin (AS) states ${\bf S} \parallel {\bf H}$. On the
AS line
there is a degeneracy in the orientation of the orbital momentum: ${\bf L}\perp
{\bf H}$. Thus a transition from the SW or BS modes to the HM-modes is signalled
by a textural transition.

We believe that HM4 represents the
canonical HM state which has similar coherent spin-superfluid properties as the
HPD. The existence of this state is verified by minimizing the
dipole and spectroscopic energies,
$F_D(l_z,s_z)- (\omega_L-\omega_{\rm RF})S_z$, where
\cite{Kharadze-Vachnadze}
$$F_D(s_z,l_z)=(\chi/10\gamma^2) \Omega_B^2 [1
+2l_z^2s_z^2+(1-l_z^2)(1-s_z^2)-$$
$$-(2/3)(1+l_z)(1+s_z)
\sqrt{(1-l_z^2)(1-s_z^2)}]\;.  \eqno(4)$$
Here $\Omega_B$ is the characteristic longitudinal resonance frequency of
$^3$He-B.  We observe the HM4 mode in the region of
relatively large negative frequency shift, where the transverse spin
$S_\perp=S\sin\beta$ is then given by
$$\sin\beta\approx \left( 1+ w^2/4 \right)^{-1/2}~,~w= 15
\omega_{\rm RF}(\omega_{\rm RF}-\omega_L)/2\Omega_B^2. $$
For $T/T_c >0.995$ and   $H -\omega_{RF}/\gamma> 0.05$ Oe one has
$\beta\ll 1$, in agreement with our observation that $\bf S$
is almost aligned along ${\bf H}$.

The other HM states at $\omega_{\rm RF}>\omega_L $ cannot be investigated
analytically, but we compare them to our numerical simulations. The results
reproduce the measured properties of the HM1 and HM3 states, such as the
very  small dispersion of HM1 in a wide range of positive frequency shifts, as
shown in Fig.~4. In all HM states the order parameter vector
${\bf {\hat n}}$ precesses with the frequency $\omega_{\rm RF}/2$ while the spin
precesses with the frequency $\omega_{\rm RF}$. The nature of the HM2 state
is not
clear. Its main features, large positive dispersion and $|S_z|=0.5$,
were not found in computer simulations in  homogeneous field. In
contrast to all other modes, HM2  was observed only at small frequency shifts,
$H-\omega_{\rm RF}/\gamma\sim \delta H$, where $\delta H$ is the remanent field
inhomogeneity (about 0.02 Oe in Fig.~2). For  $|\omega_{\rm RF}
-\omega_L| \lesssim \gamma\delta H$, different regions of the container may have
different sign of the frequency shift and the interpretation of the observed
unidentified features, marked with A in Figs.~1 and 2, is complicated.

{
\renewcommand{\baselinestretch}{1.0}
\small
\begin{figure}[!!!t]
  \begin{center}
    \leavevmode
\caption{
Dispersion ($S_x$), absorption ($S_y$) and the total spin ($S$)
normalized to $S_{\rm  eq}$,
from numerical simulation at two different $T$. The downward
field sweep was started from $\gamma H>\omega_{\rm RF}$
but new states appear only when $\gamma H<\omega_{\rm RF}$ (the latter
region is shown).  The parameters correspond roughly to the experimental
conditions: $P=0$, $H_{\rm RF}=0.02$ Oe,
$\omega_{\rm RF}/2\pi = 460$ kHz, $T_1=0.5$ s. Similar calculations for an
upward field sweep, starting from $\gamma H<\omega_{\rm RF}$, give
the HM4 state, when $\gamma H>\omega_{\rm RF}$, as in Fig~1.}
 \label{f.4}
 \end{center}
\end{figure}
}
{\it Zero-Magnetization Mode}.--The $p\approx 0$ state has
been observed in both containers at $\omega_{\rm RF}>\omega_L$ (Fig.~2).
This mode is strongly affected by dissipation and the  dipole
torque. The existence of states with $p\approx 0$ was first discussed in
Ref.~\cite{Sonin}. However, the observed ZM state has no simple
analytic description and we compare it to numerical simulations
(Fig.~4). The identification of the state is supported by measurements of the
magnitude of
$S$ with pulsed NMR. We have found that the FID amplitude decreases with
increasing
frequency shift from $p \approx 0.3$ at small frequency shifts down to about the
noise level $p\sim 0.02$ at larger shifts. The FID amplitude varies with time
in a manner which shows that ${\bf S}$ is not exactly parallel to
${\bf H}$. All these features are in agreement with computer simulations, which
also show that at large frequency shifts the transverse magnetization $S_\perp
\propto T_c-T$ does not depend on $H_{\rm RF}$. These properties were
checked and confirmed in the quartz cell.

{\it Discussion}.--There are differences in the results obtained with the two
containers. The HM states were not observed in the quartz cell while the
BS mode was not seen in the lavsan cell. This may depend on cell geometry and
dimensions,  field inhomogeneity, or relaxation at the walls. $T_1$ was measured
to be 40 s in the quartz and 1.1 s in the lavsan cells in the normal phase just
above $T_c$. In the simulations  surface relaxation is lumped into an effective
bulk $T_1$. The creation of the new states is found to be sensitive to the
choice
of $T_1$.  Most likely the differences arise from the fact that a textural
transition to the BS mode triggers a transition to the HM or ZM
states in the lavsan cell where they are more easily created  in the whole
volume
simultaneously.

{\it Conclusions}.--We have observed new precessing modes in $^3$He-B.  Some of
these  had been proposed theoretically
\cite{Smith-Brinkman,Sonin,Kharadze-Vachnadze}, but it was not known how to
obtain
them in stable state.  It is now clear  that these modes can appear only
after an initial deflection of $\bf S$  by
$\beta \approx 180^\circ$, where the HPD mode with
${\bf L}\parallel {\bf H}$ becomes unstable towards a reorientation to
${\bf L}\perp {\bf H}$.  Far from $T_c$ it is difficult to reach this region of
instability or to stay there for  the time needed to reorient ${\bf L}$. This is
prevented by the dipole torque and by Leggett-Takagi dissipation, which rapidly
grows with increasing $\beta$ above
$\theta_L$. We have  shown that favourable conditions for the new states
exist in
the vicinity of $T_c$, where $F_D \propto T_c-T$ is small, and
conventional modes with $S=S_{eq}$ can be destabilized. The
HM4 state should have the same phase coherent features
as the HPD. It will be  interesting to study its  spin-superfluid
properties and whether  it  can also be stabilized in
the limit of low $T$ where dissipation is small.

\vfill\eject

\end{document}